\documentclass[twocolumn,showpacs,preprintnumbers,amsmath,amssymb,aps,superscriptaddress,prb]{revtex4}
\usepackage{graphicx}
\usepackage{dcolumn}
\usepackage{bm}
\usepackage[bookmarks,hypertex]{hyperref}
\usepackage{multirow}
\usepackage{rotating}
\usepackage[utf8]{inputenc}

\begin{document}

\title{
Interplay between Nitrogen Dopants and Native Point Defects in Graphene
}

\author{Zhufeng Hou}
\email{hou.z.aa@m.titech.ac.jp}
\affiliation{Department of Organic and Polymeric Materials, Graduate School of Science and Engineering,
Tokyo Institute of Technology, 2-12-1 S5-20, Ookayama,Tokyo 152-8552, Japan}
\author{Xianlong Wang}
\affiliation{Department of Organic and Polymeric Materials, Graduate School of Science and Engineering,
Tokyo Institute of Technology, 2-12-1 S5-20, Ookayama,Tokyo 152-8552, Japan}
\author{Takashi Ikeda}
\affiliation{Condensed Matter Science Division, Quantum Beam Science Directorate, Japan Atomic Energy
Agency (JAEA), 1-1-1 Kouto, Sayo, Hyogo 679-5148, Japan}
\author{Kiyoyuki Terakura}
\affiliation{Research Center for Integrated Science, Japan Advanced Institute of Science and Technology
(JAIST), 1-1 Asahidai, Nomi, Ishikawa 923-1292, Japan}
\affiliation{Department of Organic and Polymeric Materials, Graduate School of Science and Engineering,
Tokyo Institute of Technology, 2-12-1 S5-20, Ookayama,Tokyo 152-8552, Japan}
\author{Masaharu Oshima}
\affiliation{Department of Applied Chemistry, The University of Tokyo, 7-3-1 Bunkyo-ku, Tokyo 113-8656, Japan}
\author{Masa-aki Kakimoto}
\affiliation{Department of Organic and Polymeric Materials, Graduate School of Science and Engineering,
Tokyo Institute of Technology, 2-12-1 S5-20, Ookayama,Tokyo 152-8552, Japan}
\author{Seizo Miyata}
\affiliation{Department of Organic and Polymeric Materials, Graduate School of Science and Engineering,
Tokyo Institute of Technology, 2-12-1 S5-20, Ookayama,Tokyo 152-8552, Japan}
\begin{abstract}
To understand the interaction between nitrogen dopants and native point defects in graphene, we have studied the energetic stability of N-doped graphene with vacancies and Stone-Wales (SW) defect by performing the density functional theory calculations. Our results show that N substitution energetically prefers to occur at the carbon atoms near the defects, especially for those sites with larger bond shortening, indicating that the defect-induced strain plays an important role in the stability of N dopants in defective graphene. In the presence of monovacancy, the most stable position for N dopant is the pyridinelike configuration, while for other point defects studied (SW defect and divacancies) N prefers a site in the pentagonal ring. The effect of native point defects on N dopants is quite strong: While the N doping is endothermic in defect-free graphene, it becomes exothermic for defective graphene. Our results imply that the native point defect and N dopant attract each other, i.e., cooperative effect, which means that substitutional N dopants would increase the probability of point defect generation and vice versa. Our findings are supported by recent experimental studies on the N doping of graphene. Furthermore we point out possibilities of aggregation of multiple N dopants near native point defects. Finally we make brief comments on the effect of Fe adsorption on the stability of N dopant aggregation.

\end{abstract}
\maketitle
\section{Introduction}
\label{sec:1}

Graphene is a carbon allotrope with a two dimensional (2D) honeycomb lattice. Since its first successful isolation in 2004,~\cite{Novoselov04} graphene has attracted immense attention because of its 2D crystal lattice with atomic thickness and unique electronic structures.~\cite{Novoselov04,Neto09}
It has opened up exciting opportunities for developing nanoelectronic devices.~\cite{Geim07} In low dimensional systems, the chemical and physical properties of materials can be heavily affected by the lattice imperfection due to the structural defects.~\cite{Banhart2010} The point defects in graphene can be introduced via the formation of vacancies or the atomic rearrangement (e.g., Stone-Wales defect consisting of a pentagon-heptagon (5-7) pair).~\cite{Lusk2008,Banhart2010,Appelhans2010} Vacancies in graphene may be formed in low concentrations during the growth process. Alternatively, they may be created intentionally by irradiating materials with electrons or ions or by chemical treatments.~\cite{Kholmanov2011} Recent advances in microscope technologies enable us to observe the structural defects in graphene at an atomic resolution by the transmission electron microscope (TEM)~\cite{Meyer2008,Hashimoto2004,Kotakoski2011} and the scanning tunneling microscope (STM).~\cite{Ugeda2010,Kondo2010,Lahiri2010} On the other hand, recent density functional theory (DFT) calculations show that some of structural defects can induce localized levels close to the Fermi level ($E_{\mathrm{F}}$), leading to local charging~\cite{Carlsson2006} and/or local magnetic moments.~\cite{Lehtinen2004} Therefore, defects are expected to play key roles in the chemical functionality and electronic transport properties of graphene-based materials.

One simple approach to further tailor the electronic properties of graphene is the incorporation of heteroatoms. For instance, substitution of carbon with nitrogen or boron atoms injects electron or hole carriers, respectively. Such carbon-based materials containing some different elements, called carbon alloys~\cite{Tanabe2000}, have been exploited recently to be a high-promising candidate to replace a Pt-based catalyst in the polymer electrolyte fuel cell (PEFC).~\cite{Lefevre03042009,biddinger2010,Ozaki2010,Nabae2010,Wu22042011,Proietti2011} N doping into graphene can be performed either directly during synthesis or by post-synthetic treatment.~\cite{Wang2009,Lixl2009,Panchakarla2009,Zhang201004110,wei2009,Qu2010,Reddy2010,Guo2010,lin133110,Jafri2010,Wang2010acs,Deng2011cm,Sun2010,Imamura2011} Even the controllable N doping has been realized by NH$_3$ annealing after ion irradiation~\cite{Guo2010} or by NH$_3$ plasma exposure.~\cite{lin133110} The presence of large amount of defects in irradiated or plasma-treated graphene has been revealed by a pronounced \emph{D} band in the Raman spectra.~\cite{Guo2010,lin133110} After N doping, the \emph{D} band becomes more prominent, indicating that N doping is likely to induce non-negligible amount of defects and bond disorders.~\cite{Guo2010,lin133110,Zhang201004110,Deng2011cm,Imamura2011} These suggest that some mutual effects of structural defects and N doping exist during the incorporation of N into graphene. Previous DFT calculations show that the localized edge states play an important role in the stability of N substitution in graphene nanoribbons (GNRs)~\cite{Yu2008537,Liyf2009} and clusters.~\cite{Huang09} Thus, it raises a question whether the defect-induced localized states also have significant impacts on the N doping of graphene or not. Indeed, some specific configurations of substitutional N next to vacancies in GNRs have already been studied theoretically in literature.~\cite{Liyf2009,Fujimoto2011} In particular, recent work~\cite{Fujimoto2011,comment2011} deals with N
doping at monovacancy (MV) and divacancy (DV). In the present work, we
study more extensively and systematically the interplay between
structural defects and N doping using the DFT calculations with a
significantly larger supercell as described in detail below.  Furthermore, we discuss briefly the effect of Fe adsorption on the stability of N dopant arrangement, considering that the FeN$_x$ complex in graphene may play important roles in the oxygen reduction reaction in the carbon based fuel cell catalyst.~\cite{Lefevre03042009,Wu22042011,Proietti2011,DHLee2011}  We focus
on the structural and energetics aspects in this paper. The detailed
discussion on the electronic structures will be given in a separate
paper.

The remainder of this paper is organized as follows. In Sec. \ref{sec:2}, we introduce the computational methods for the calculations of N-doped graphene (N-graphene). The computed interaction energies between N dopants and structural defects are presented in Sec. \ref{sec:3}. Finally, we draw conclusions in Sec. \ref{sec:4}.

\section{Method and computational details}
\label{sec:2}
DFT calculations are performed with the \texttt{PWSCF} code of the Quantum \texttt{ESPRESSO} suite~\cite{Baroni} in a plane-wave ultrasoft-pseudopotential~\cite{Rappe90} approach. The exchange-correlation functional is treated by the generalized gradient approximation (GGA) after Perdew, Burke, and Ernzerhof.~\cite{Perdew96} Spin polarization is taken into account if it exists. The kinetic energy cutoffs for wave function and charge are set to 35 and 350 Ry, respectively. A supercell constructed by the 9$\times$9 extension of the hexagonal unit cell of graphene with the calculated lattice constant of 2.463 \AA~for perfect graphene is employed to study the structural defects and the substitution of carbon by nitrogen. To avoid the spurious interaction between graphene layers, the vacuum thickness in the supercell is set to 12.0 \AA. A $3\times3\times1$ \emph{k}-point grid in the Monkhorst-Pack scheme~\cite{Monkhorst1976} is employed to sample the Brillouin zone (BZ) of the above supercell. During geometry optimization, all atoms are fully relaxed until residual forces on constituent atoms are smaller than 0.01 eV/\AA.

\begin{figure*}[htbp!]
\begin{center}
\begin{minipage}[t]{12.5cm}
\includegraphics*[width=12.5cm]{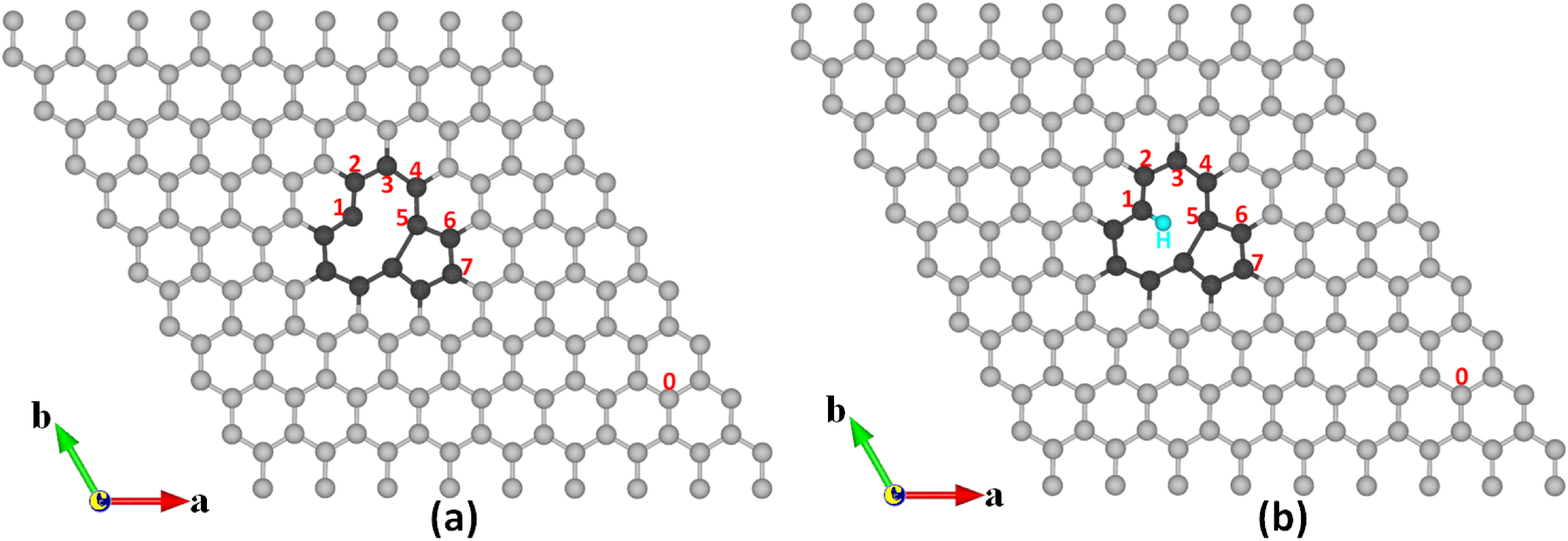}
\end{minipage}
\begin{minipage}[t]{12.5cm}
\includegraphics*[width=12.5cm]{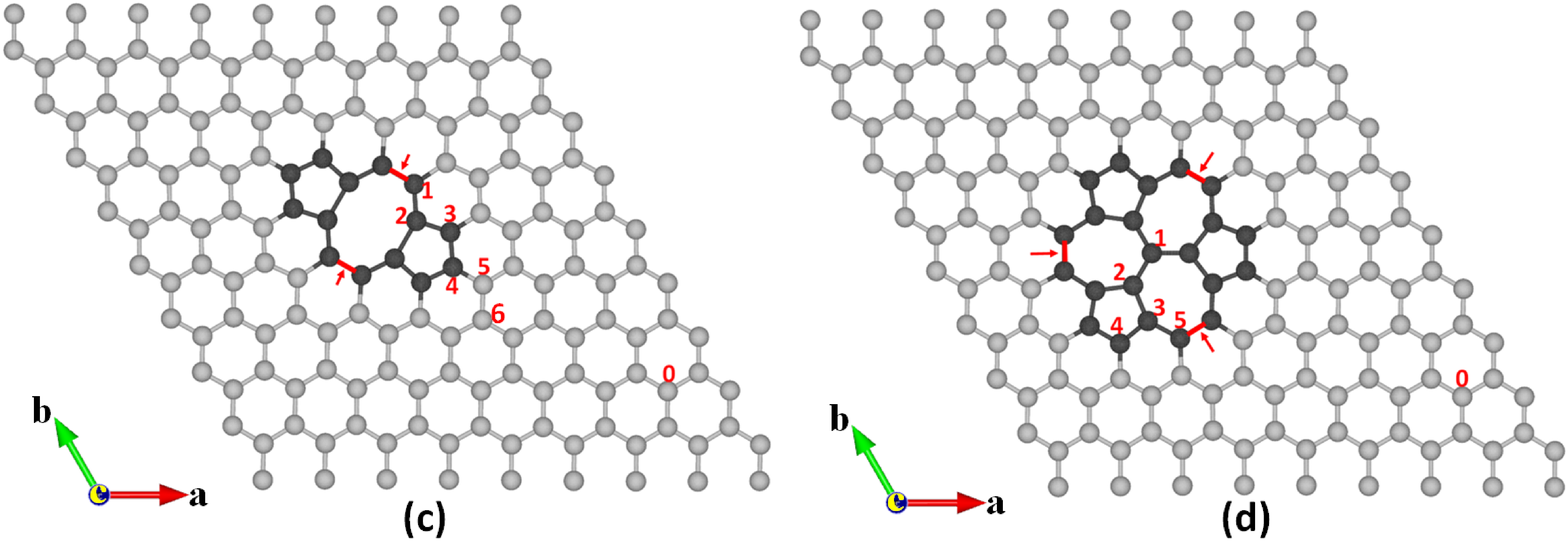}
\end{minipage}
\begin{minipage}[t]{12.5cm}
\includegraphics*[width=12.5cm]{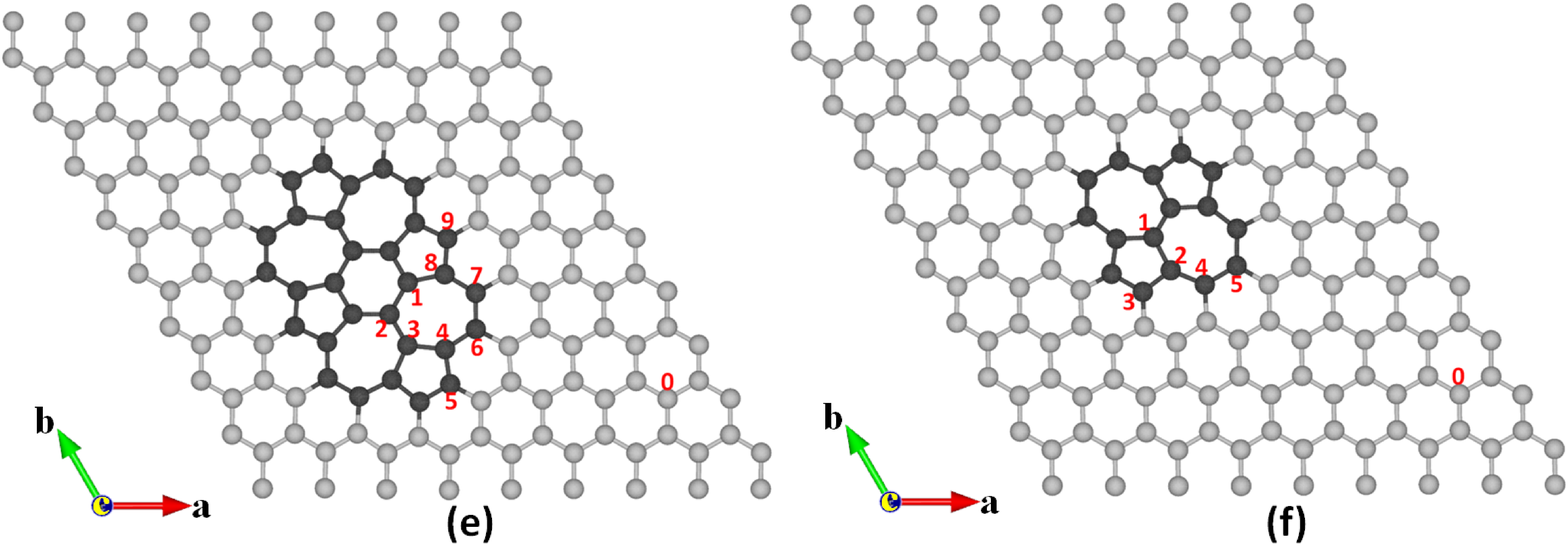}
\end{minipage}
\caption{\label{fig:1} (Color online) The optimized atomic structures for defective graphene with (a) MV, (b) H-MV, (c) 5-8-5 configuration of DV (denoted as 5-8-5 DV), (d) 555-777 configuration of DV (denoted as 555-777 DV), (e) 5555-6-7777 configuration of DV (denoted as 5555-6-7777 DV), and (f) SW defect. C atoms in defect region are represented by black balls and other C atoms are represented by gray ones. The inequivalent sites for C atom in the defect region are labeled by nonzero numbers, while C atom labeled with 0 stands for the site far from the defect region. The arrows in (c) and (d) indicate possible rotated bonds for the reconstruction.}
\end{center}
\end{figure*}

In this work, the most common types of native point defects (NPDs) are studied as representatives of structural defects in graphene, namely, MV, DV, and Stone-Wales (SW) defect. Additionally, the hydrogenated monovacancy (H-MV) and the reconstructed DVs are also studied. The optimized atomic structures of defective graphene are shown in Fig.~\ref{fig:1}. For DVs, besides the normal 5-8-5 configuration [5-8-5 DV, Fig. \ref{fig:1}(c)], we consider two variants. The 555-777 configuration [555-777 DV, Fig. \ref{fig:1}(d)] can be reconstructed from the 5-8-5 DV by in-plane 90$^{\circ}$ rotation of either bond indicated by arrows shown in Fig. \ref{fig:1}(c). Similarly, the 5555-6-7777 configuration [5555-6-7777 DV, Fig. \ref{fig:1}(e)] can be reconstructed from the 555-777 DV by rotating one of the three bonds indicated by arrows in Fig. \ref{fig:1}(d). For N doping, we consider all the single N substitution of inequivalent carbons $\mathrm{C}i$ in the defect region as marked in Fig.~\ref{fig:1} in addition to carbon $\mathrm{C}0$ far from the defect for each defective graphene. Here, such a doped N is denoted as N$_{\mathrm{C}i}$ according to Kr\"{o}ger and Vink.~\cite{Kroger56}

\section{Results and discussion}
\label{sec:3}

\subsection{Interaction between two substitutional N atoms in perfect graphene}
\label{sec:3.1.2}
\begin{figure*}[htbp!]
\begin{center}
\begin{minipage}[t]{11.0cm}
\includegraphics*[width=11.0cm]{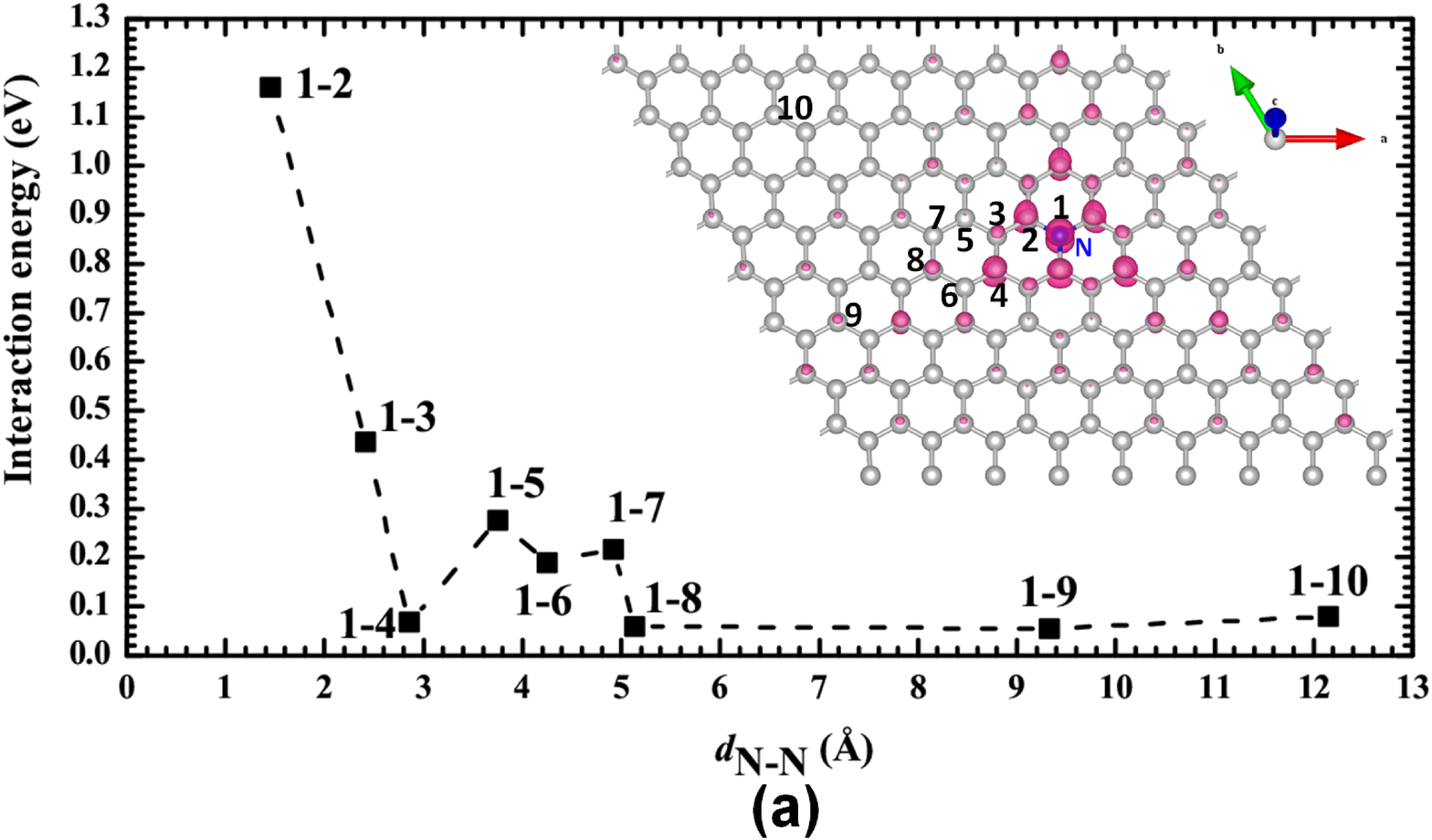}
\end{minipage}
\begin{minipage}[t]{4.5cm}
\includegraphics*[width=4.5cm]{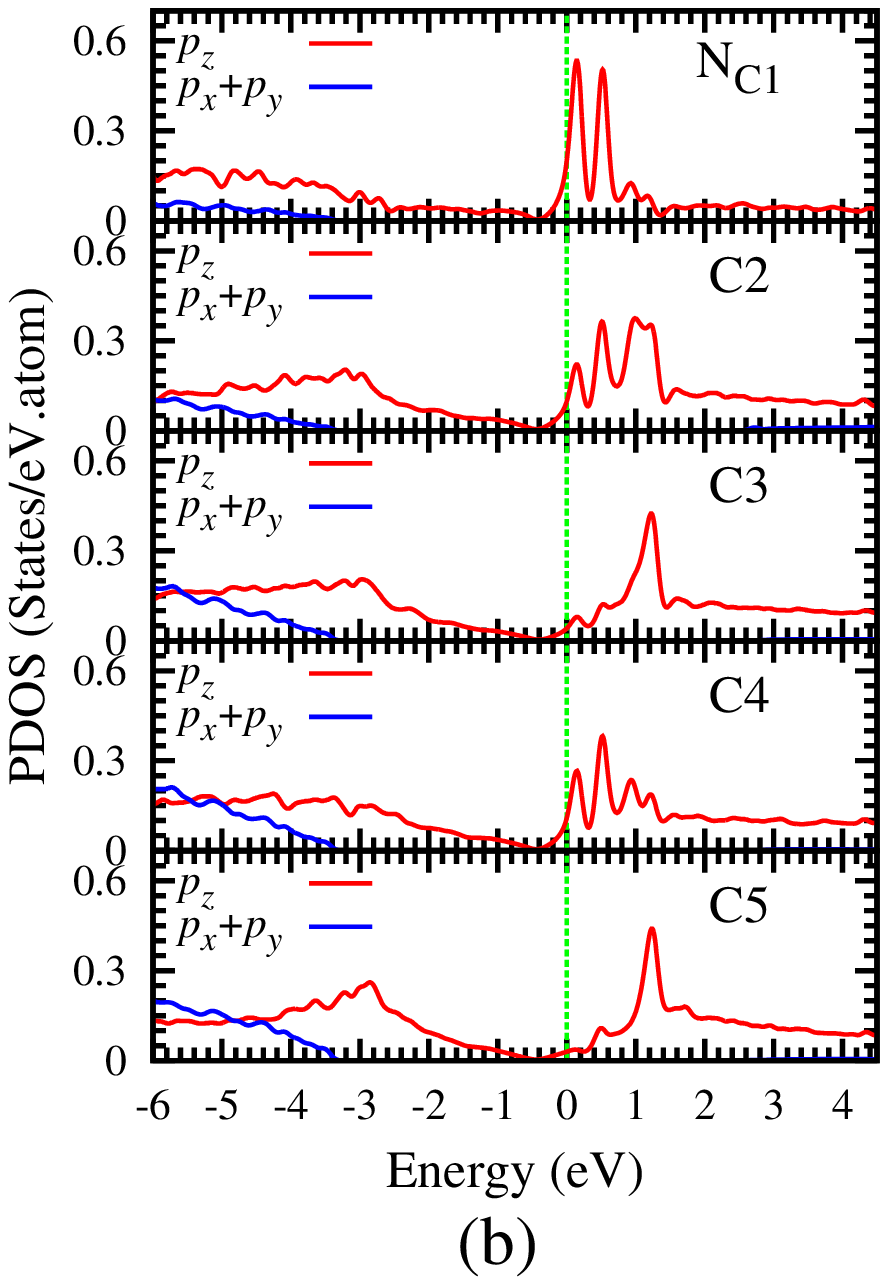}
\end{minipage}
\caption{\label{fig:2} (Color online) (a) The interaction energy between two doped N atoms in defect-free graphene. The inset shows local density of states (0.003 \emph{e}/\AA$^3$) integrated from Fermi level ($E_\mathrm{F}$) to $E_\mathrm{F} +0.25 $ eV for single N dopant in defect-free graphene and the sites considered for the N substitution. Here $E_\mathrm{F}$ is the Fermi level of defect-free graphene with single N dopant. (b) The projected density of states (PDOS) for the \textit{p} orbitals of C and N atoms in defect-free graphene with single N dopant.}
\end{center}
\end{figure*}

As a first step to examine the aggregation of substitutional nitrogens, the interaction energy $\Delta E_\mathrm{N,N}$ between two substitutional N atoms is calculated according to
\begin{equation}\label{eq:1}
\Delta E_\mathrm{N,N} = E_\mathrm{2N}+ E_{0}-2E_\mathrm{N},
\end{equation}
where $E_{0}$, $E_\mathrm{N}$, and $E_\mathrm{2N}$ are the total energies of the supercells of graphene containing zero, one, and two substitutional N atoms, respectively. Figure~\ref{fig:2}(a) shows our computed $\Delta E_\mathrm{N,N}$ for two doped nitrogens in the defect-free graphene as a function of their distance. The interaction energy is generally positive (i.e., repulsive) and is found to be much higher for near N-N dimers than for distant ones. For example, as shown in Fig. \ref{fig:2}(a), the interaction energy of the second nearest N$_\mathrm{C1}$-N$_\mathrm{C3}$ pair is about 0.36 eV higher than that of the well-separated N$_\mathrm{C1}$-N$_\mathrm{C10}$ pair. Thus, such a configuration of two doped N atoms at the second nearest neighbors is unlikely due to its low stability, contrary to the recent proposal based on the STM images of N-graphene.~\cite{Deng2011cm} It is interesting to note that two doped N atoms at the third or seventh nearest neighbors (i.e., N$_\mathrm{C1}$-N$_\mathrm{C4}$ or N$_\mathrm{C1}$-N$_\mathrm{C8}$ pairs, respectively) have very small interaction energy. In a recent study,~\cite{Xiang2011} the stabilities of N$_\mathrm{C1}$-N$_\mathrm{C4}$ and N$_\mathrm{C1}$-N$_\mathrm{C8}$ pairs are ascribed to the low Coulomb repulsion due to the anisotropic electron charge density redistribution induced by the N substitution. Here we give a different interpretation to the trend seen in Fig.~\ref{fig:2}(a).

The quantity in the inset of Fig.~\ref{fig:2}(a) is obtained by integrating the local density of states from $E_\mathrm{F}$ to $E_\mathrm{F} +0.25 $ eV in the presence of a single N dopant at the C1 site in defect-free graphene. As the impurity resonance state appears just above $E_\mathrm{F}$ as shown in the density of states in Fig.~\ref{fig:2}(b), the above quantity corresponds to the weight distribution of the tail of the impurity state. Therefore, if the second N is doped at the site where the tail has a large weight, the two N atoms can form a stronger bond. Therefore we expect that C2, C4, and C8 sites are favorable for the second N dopant. However the N$_\mathrm{C1}$-N$_\mathrm{C2}$ pair has too strong of a core-core repulsive interaction due to the shortest N-N distance. As the fourth, fifth, and sixth nearest neighbors (i.e., C5, C6, and C7 atoms) of N$_\mathrm{C1}$ have less weight of impurity state tail, the N$_\mathrm{C1}$-N$_\mathrm{C5}$, N$_\mathrm{C1}$-N$_\mathrm{C6}$, and N$_\mathrm{C1}$-N$_\mathrm{C7}$ pairs have lower stability than the N$_\mathrm{C1}$-N$_\mathrm{C4}$ pair. More detailed discussion on the electronic
structures of N-N pairs will be given in a separate paper. Finally we point out that two H adatoms are stably adsorbed on graphene in a way~\cite{Ferro2008} similar to a N-N pair treated here. The stability of two H adatoms was explained in terms of mesomeric effect, which is just another way to take into account the structure of the wave function at the Fermi level.~\cite{Ferro2008}

\subsection{Formation energy of native point defect in undoped graphene}
Before discussing the N doping of graphene with NPDs, we assess the stability of NPDs in graphene by calculating their formation energy $\Delta E_0(\mathrm{d})$ as
\begin{equation}\label{eq:2}
\Delta E_0(\mathrm{d}) = E_\mathrm{d}- E_\mathrm{0}+n_\mathrm{C}\mu_\mathrm{C}-n_\mathrm{H}\mu_\mathrm{H},
\end{equation}
where $E_\mathrm{d}$ and $E_\mathrm{0}$ are the total energies of a supercell of undoped graphene with and without a NPD. $\mu_\mathrm{C}$ and $\mu_\mathrm{H}$ are the respective chemical potentials of C and H, which are taken as the total energy per C atom of graphene for C and half of the total energy of an isolated H$_2$ molecule for H. $n_\mathrm{C}$ ($n_\mathrm{H}$) is the number of C (H) atoms removed (attached) when NPD is formed. Our computed formation energies are summarized along with those found in literature in Table~\ref{tab:1}. For MV, H termination is found to reduce the formation energy by 1.948 eV/per H, indicating that H-MV is energetically much more favorable. On the other hand, for DVs, we find, in agreement with previous studies,~\cite{Barbary2003,Lee2005,Cretu2010,Banhart2010,Appelhans2010} that the most stable configuration is the reconstructed one with three pentagons and three heptagons (denoted as 555-777), rather than the original 5-8-5 DV. It is also seen that the formation energy of SW defect is a few eV lower than that of MV and DVs, which is consistent with the results reported in Ref.~\onlinecite{Carlsson2006}. All of these validate the computational setup chosen in the present study.

\begin{table*}[htbp]
  \caption{Formation energy [$\Delta E_0(\mathrm{d})$, in eV] of MV, H-MV, DVs, and SW defect in undoped graphene. }
  \label{tab:1}
  \begin{tabular}{lc ccccc}
    \hline
    Defect type & MV  & H-MV  &  \multicolumn{3}{c}{DV} & SW\\
    \cline{4-6}
    Configuration & 5-9 & 5-9 & 5-8-5  & 555-777 &  5555-6-7777 & 55-77\\
    \hline
     $\Delta E_0(\mathrm{d})$ (This work) & 7.536 &  5.588  & 7.442    &  6.616  &  6.963  & 4.875\\
       $\Delta E_0(\mathrm{d})$ (Ref.~\onlinecite{Banhart2010})   & 7.3-7.5 &        & 7.2-7.9 & 6.4-7.5  & 7.0      & 4.5-5.3 \\
    \hline
  \end{tabular}
\end{table*}

\subsection{Interaction between a single N dopant and a native point defect}
\label{sec:3.2}

The stable positions of N substitution in defective graphene are explored by considering all inequivalent sites around the defect as shown in Fig.~\ref{fig:1} for each type of NPDs. The interaction energy $\Delta E_\mathrm{N,d}$ between a N dopant and a NPD is defined as follows:
\begin{eqnarray}\label{eq:3}
 \Delta E_\mathrm{N,d} &=&  (E_\mathrm{N+d}+E_0)-(E_\mathrm{N}+E_\mathrm{d}),
\end{eqnarray}
where $E_\mathrm{N+d}$, $E_0$, $E_\mathrm{N}$, and $E_\mathrm{d}$ are the total energies of the supercells for the N-graphene with NPDs, the perfect graphene, the N-graphene, and the defective graphene, respectively.  The negative value of $\Delta E_\mathrm{N,d}$ indicates the attractive interaction between N dopant and NPD.  Equation~(\ref{eq:3}) can also be expressed in the following way.  The formation energy of an N dopant in a defective graphene $\Delta E_\mathrm{d}(\mathrm{N})$ and the one in a perfect graphene $\Delta E_0(\mathrm{N})$ are given by
\begin{eqnarray}\label{eq:4-5}
 \Delta E_\mathrm{d}(\mathrm{N})&=&E_\mathrm{N+d}-E_\mathrm{d}+\mu_\mathrm{C}-\mu_\mathrm{N}, \\
 \Delta E_0(\mathrm{N})&=&E_\mathrm{N}-E_0+\mu_\mathrm{C}-\mu_\mathrm{N}.
\end{eqnarray}
Then we obtain
\begin{equation}\label{eq:6}
\Delta E_\mathrm{N,d} = \Delta E_\mathrm{d}(\mathrm{N}) - \Delta E_{0}(\mathrm{N}).
\end{equation}
The right-hand side of this equation is the difference in the N dopant formation energy between a defective graphene and a perfect one.

\begin{figure*}[htbp!]
\begin{center}
\begin{minipage}[t]{8.0cm}
\includegraphics*[width=8.0cm]{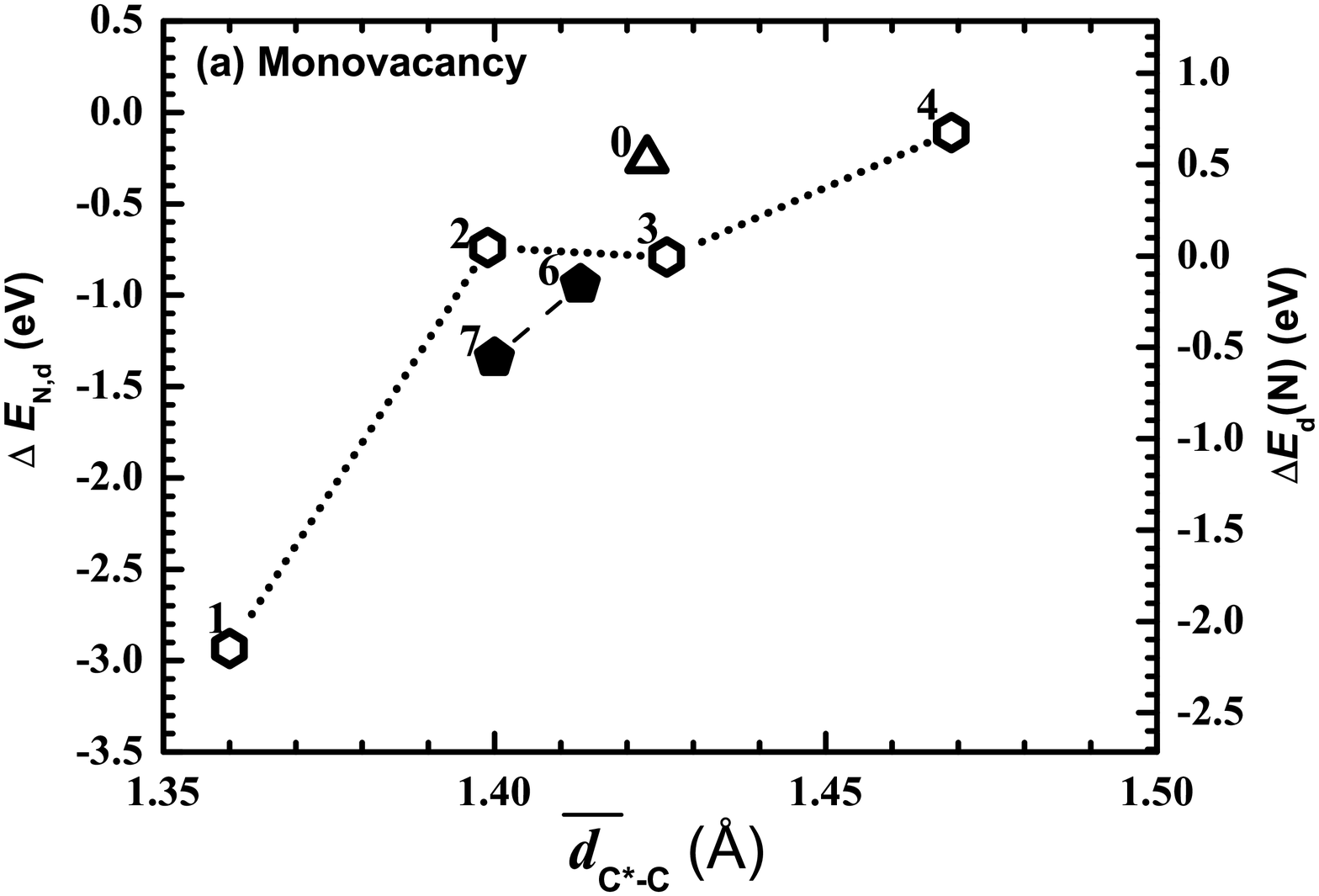}
\end{minipage}
\begin{minipage}[t]{8.0cm}
\includegraphics*[width=8.0cm]{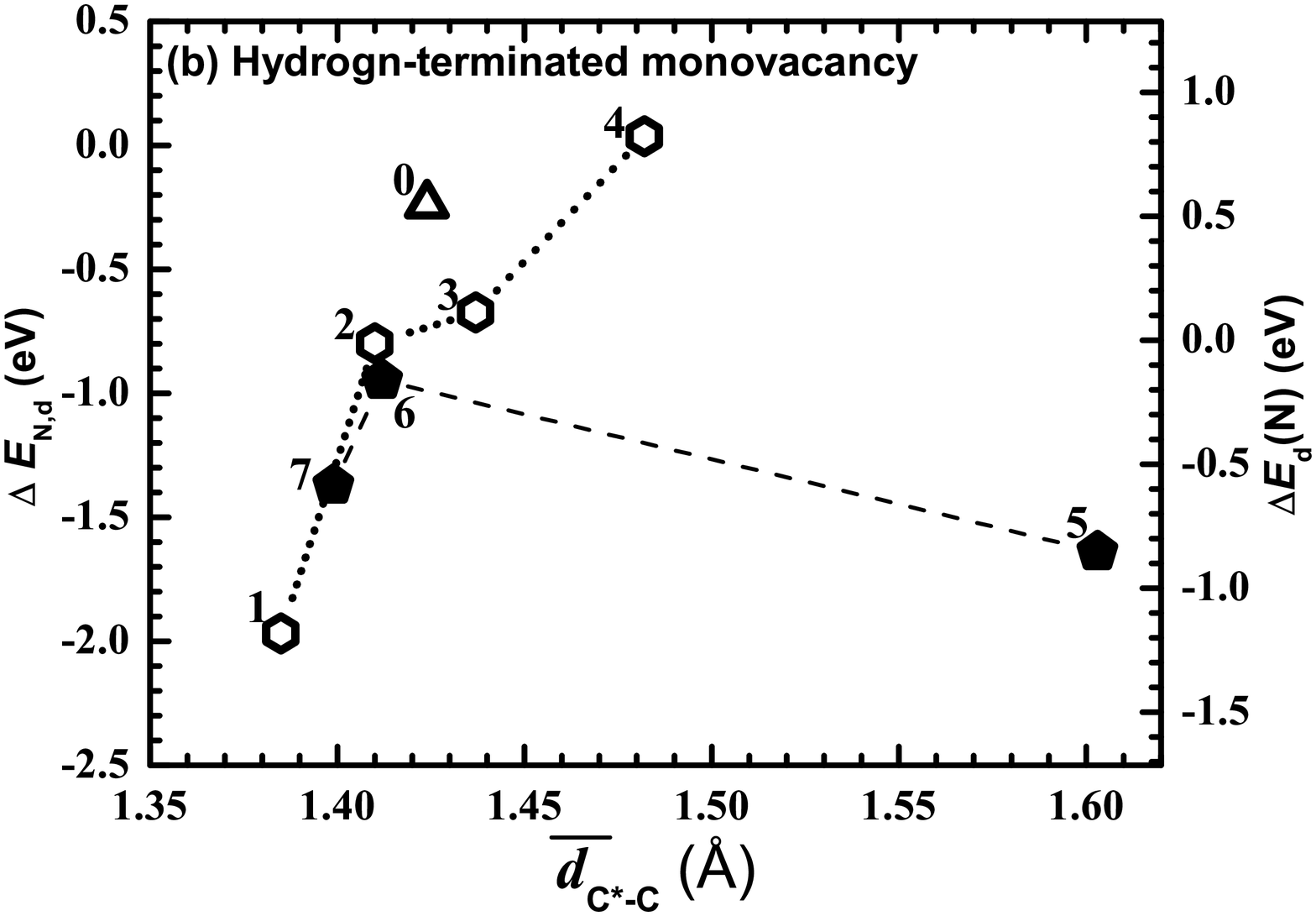}
\end{minipage}
\begin{minipage}[t]{8.0cm}
\includegraphics*[width=8.0cm]{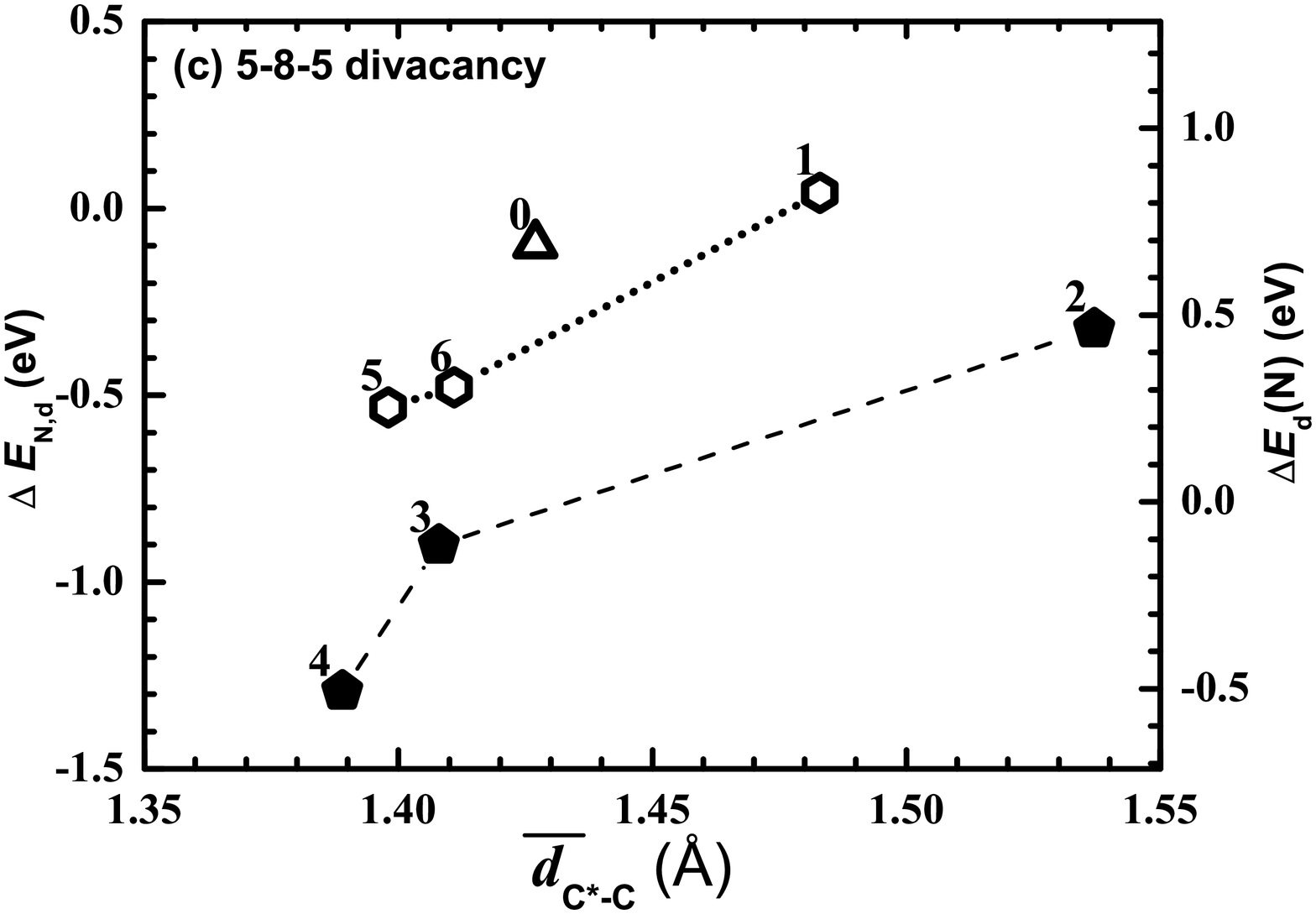}
\end{minipage}
\begin{minipage}[t]{8.0cm}
\includegraphics*[width=8.0cm]{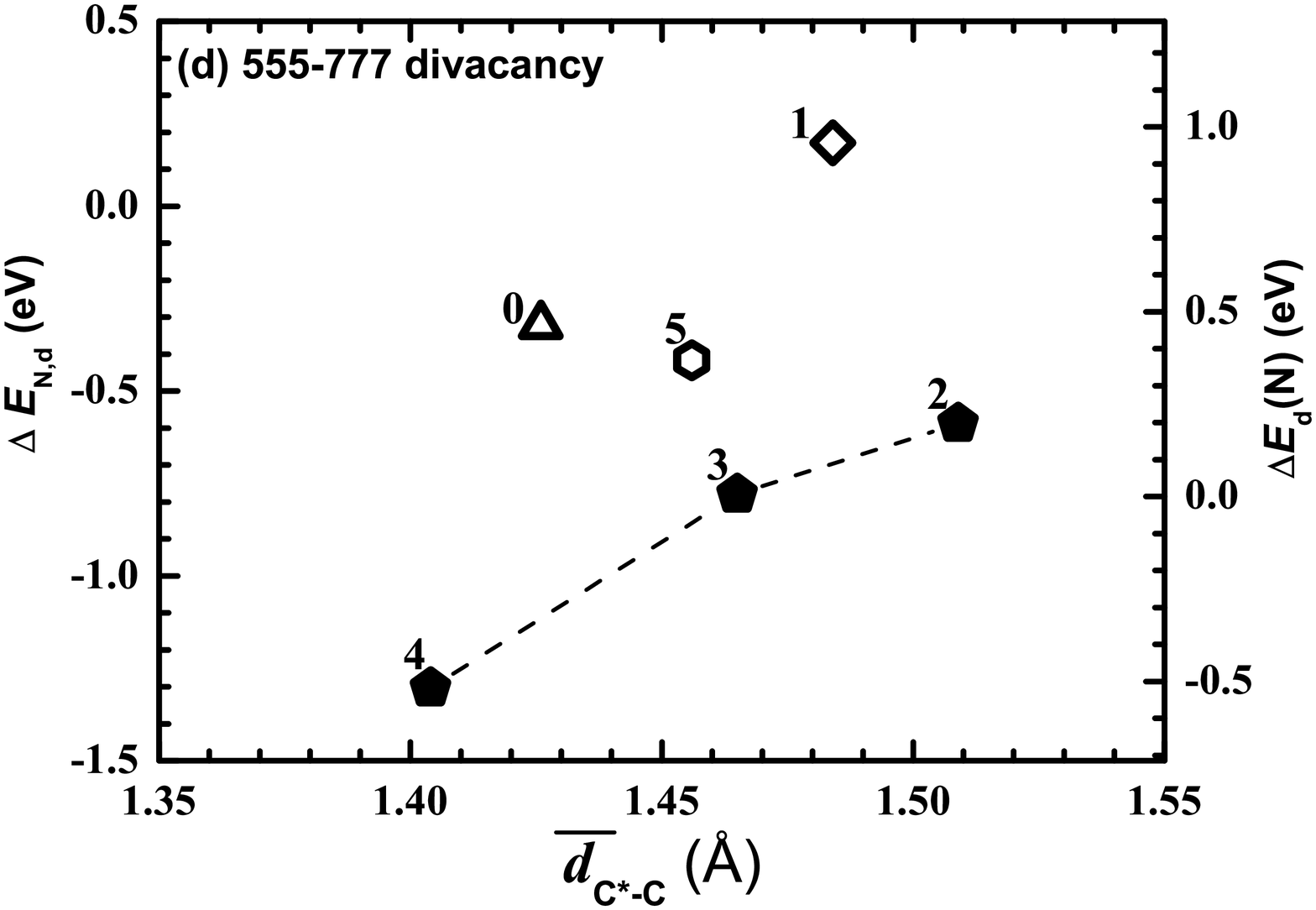}
\end{minipage}
\begin{minipage}[t]{8.0cm}
\includegraphics*[width=8.0cm]{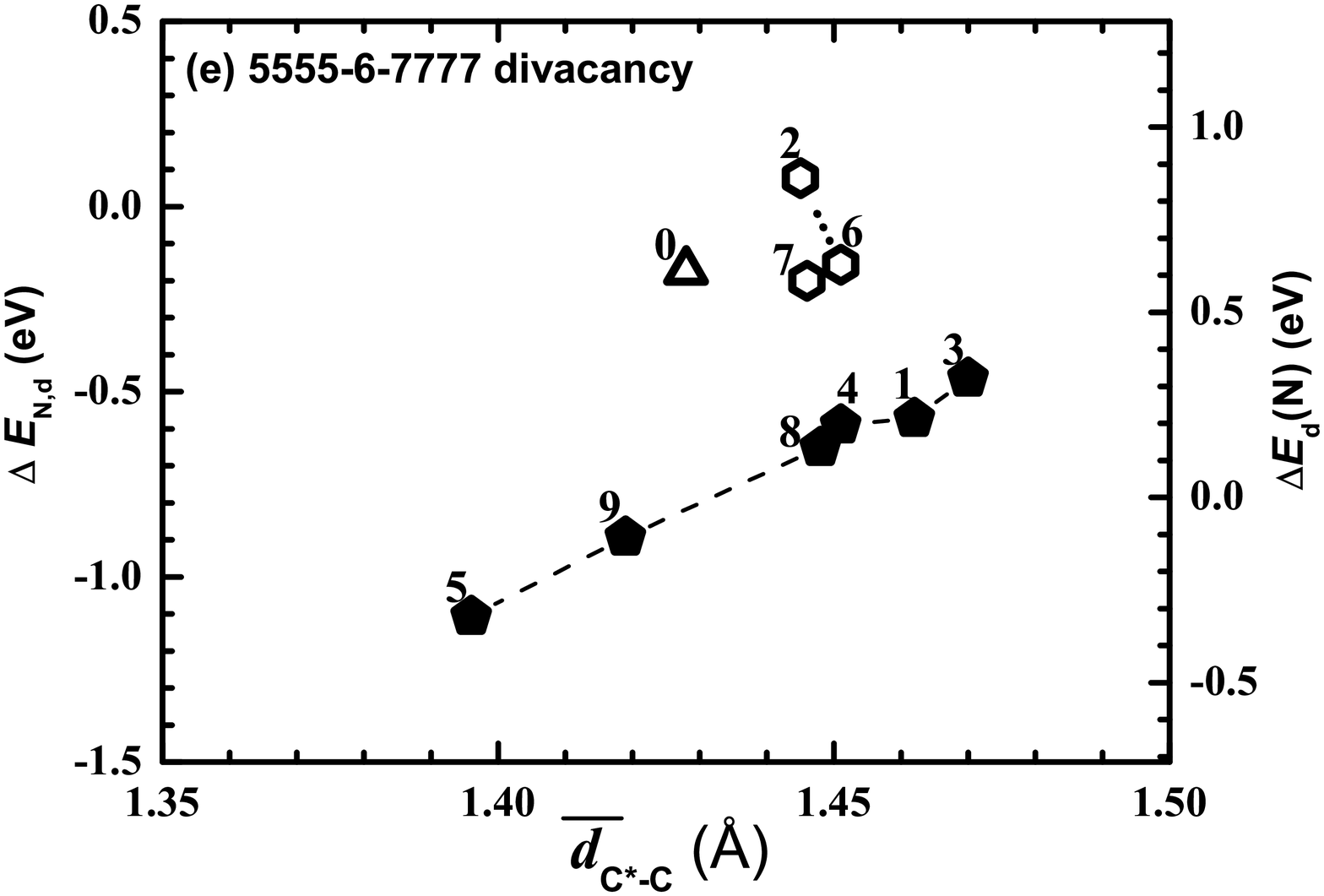}
\end{minipage}
\begin{minipage}[t]{8.0cm}
\includegraphics*[width=8.0cm]{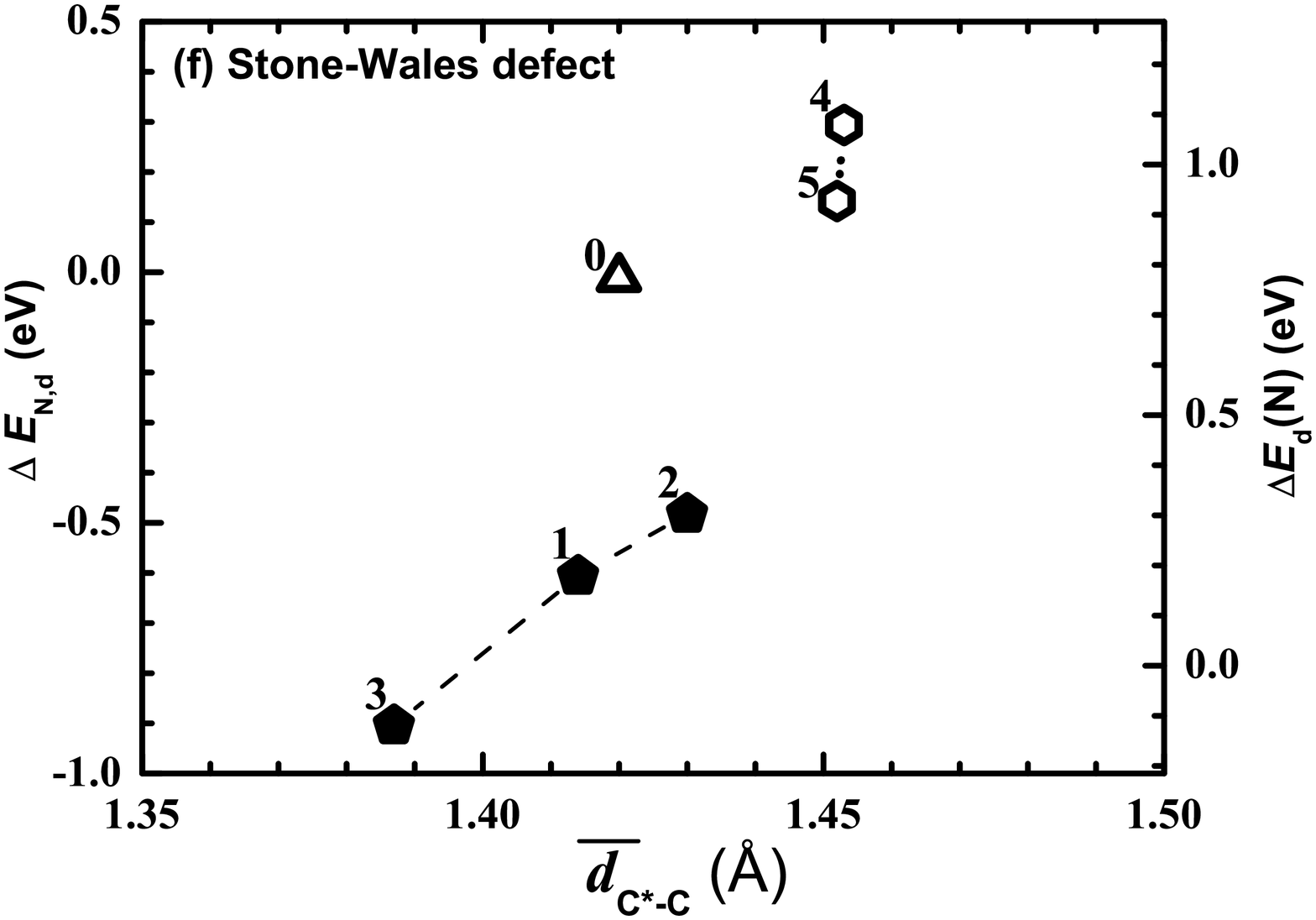}
\end{minipage}
\caption{\label{fig:3} The interaction energy ($\Delta E_\mathrm{N,d}$, left $y$ axis) between single N dopant ($m=1$) and NPD and the formation energy of N substitution [$\Delta E_\mathrm{d}(\mathrm{N})$, right $y$ axis] in defective graphene with (a) MV, (b) H-MV, (c) 5-8-5 DV, (d) 555-777 DV, (e) 5555-6-7777 DV, and (f) SW defect as function of average C$^{\ast}$-C bond length ($d_\mathrm{C^{\ast}-C}$) around the substitution site (C$^{\ast}$) before N doping. The open hexagonal symbols stands for C$^{\ast}$ at a six-membered ring, the solid pentagonal ones for C$^{\ast}$ at a five-membered ring, and the up-triangle symbols for C0$^{\ast}$ far away from the defect region. The positions of substitution sites $\mathrm{C}i^{\ast}$ (here $i$ is the number used to mark the inequivalent sites around the defect region) can be referred to Fig.~\ref{fig:1}.}
\end{center}
\end{figure*}

The results of the calculation for $\Delta E_\mathrm{N,d}$ and $\Delta E_\mathrm{d}(\mathrm{N})$ are shown in Fig.~\ref{fig:3} for a single N dopant in various defective graphenes as a function of the average bond length ($\bar{d}_\mathrm{C^{\ast}-C}$) around the substitution site (C$^{\ast}$). $\Delta E_0(\mathrm{N})$ is estimated as 0.785 eV. For each type of NPDs, the C0 site (Fig.~\ref{fig:1}) is separated from the NPD by several C-C bonds and can be an approximated site in a perfect graphene.  Therefore, $\Delta E_\mathrm{N,d}$ for the C0 site is nearly zero (at most about -0.3 eV for the MV case) in Fig.~\ref{fig:3}.  We note the following facts in Fig.~\ref{fig:3}:

(1) For all the stable dopant configurations, $\Delta E_\mathrm{N,d}$ is negatively large implying that a N dopant and a NPD attract each other and that the N dopant formation energy is significantly reduced to being even exothermic by the presence of NPD.

(2) For the most stable site of N dopant in each of the defective graphene, the bond (C$^{\ast}$-C) (before N doping) associated with the substitution site (denoted as C$^{\ast}$)  is the shortest and much shorter than the corresponding C-C bond (1.422 \AA) in the perfect graphene, suggesting that the N dopant tends to be located at the site with larger compressed strain in defective graphene.

(3) In the presence of MV, the most stable site for the N dopant is the C1 site, whose configuration may be called pyridinelike with a dangling $\sigma$ state,~\cite{Lehtinen2004} and moreover, even if the dangling $\sigma$ state is passivated by H as in H-MV,~\cite{Lehtinen2004} the N dopant still energetically prefers the C1 site.

(4) For DVs and SW defect, the N dopant energetically prefers the vertex sites of the five-membered rings, the trend being observed in the N-doped carbon conjugated materials,~\cite{Gao2010} which is ascribed to higher aromaticity according to H\"{u}ckel's rule.

Most importantly our results suggest that the creation of defects in graphene before introducing N will enhance the incorporation of N into graphene. This would support the recent experimental studies, in which the authors reported that the NH$_3$ annealing of graphene after N$^{+}$-ion irradiation~\cite{Guo2010} or the NH$_3$ plasma exposure~\cite{lin133110} of graphene can be used to realize the N doping of graphene in a controllable manner.

So far, $\Delta E_\mathrm{N,d}$ has been discussed from the view point of the effect of NPD on the N dopant formation energy.  Similarly, $\Delta E_\mathrm{N,d}$ can be regarded as the difference in the formation energy of defects after and before N doping.  More concretely, the NPD formation energy in the presence of N dopant $\Delta E_\mathrm{N}(\mathrm{d})$ is given by
\begin{equation}\label{eq:6,7}
 \Delta E_\mathrm{N}(\mathrm{d}) = E_\mathrm{N+d}-E_\mathrm{N}+n_\mathrm{C}\mu_\mathrm{C}-n_\mathrm{H}\mu_\mathrm{H}.
\end{equation}
The one in a perfect graphene $\Delta E_0(\mathrm{d})$ is given by Eq.~(\ref{eq:2}). Then we obtain
\begin{equation}\label{eq:8}
\Delta E_\mathrm{N,d} = \Delta E_\mathrm{N}(\mathrm{d}) - \Delta E_{0}(\mathrm{d}).
\end{equation}

The results depicted in Fig.~\ref{fig:3} can now be regarded as the energy gain in the NPD formation energy caused by the presence of N dopant.  Therefore, they suggest that N doping would increase the probability of point defect generation in graphene. In experimental studies, the Raman spectra of N-graphene have an intense \emph{D} band, which is generally ascribed to the defects in the $sp^2$ C network of graphene.~\cite{Guo2010,lin133110,Zhang201004110,Deng2011cm}

\subsection{Multiple N dopants near native point defects}
\label{sec:3.2.5}

\begin{figure*}[htbp]
\begin{center}
\begin{minipage}[t]{13.0cm}
\includegraphics*[width=13.0cm]{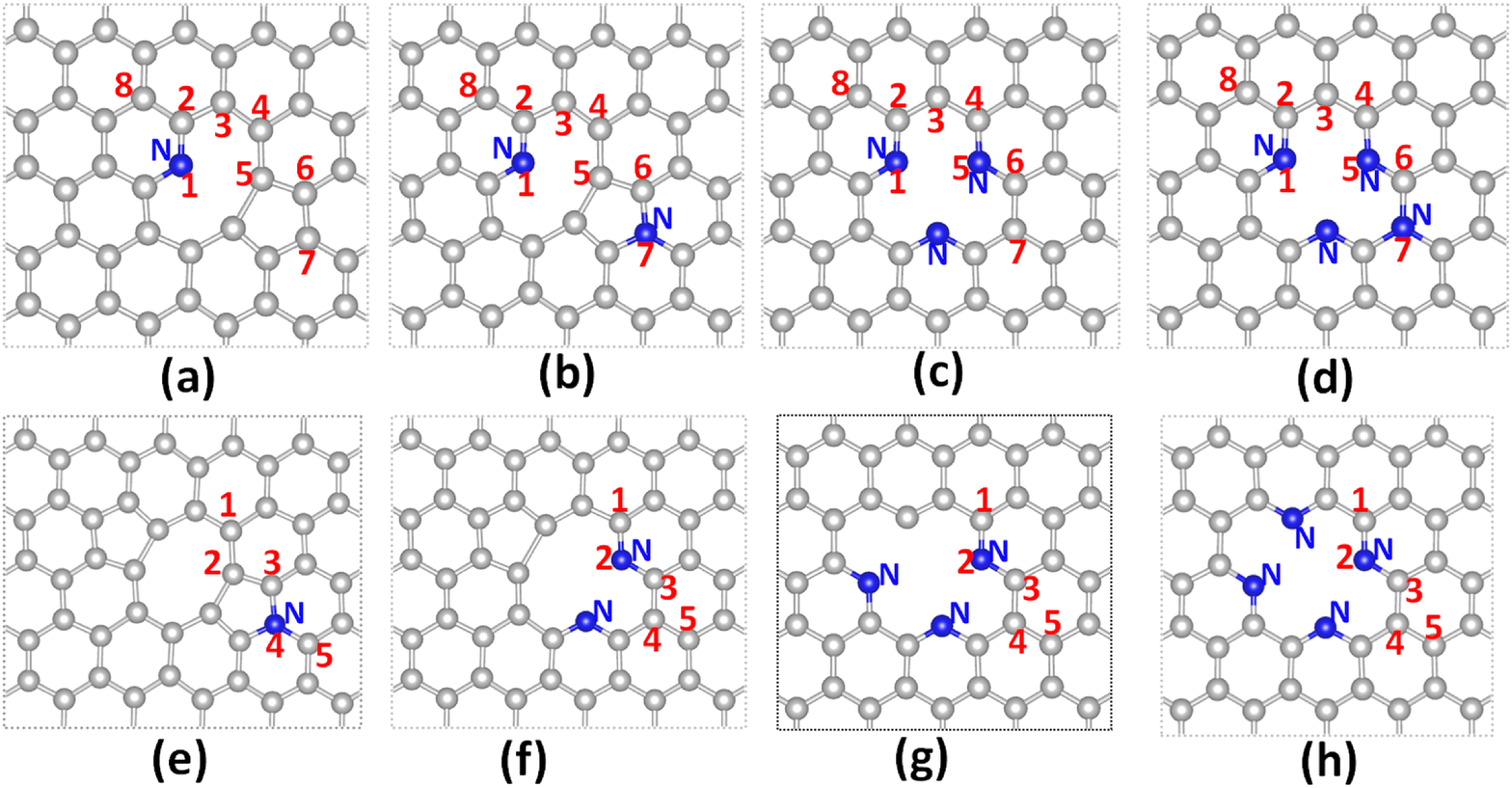}
\end{minipage}
\begin{minipage}[t]{13.0cm}
\includegraphics*[width=13.0cm]{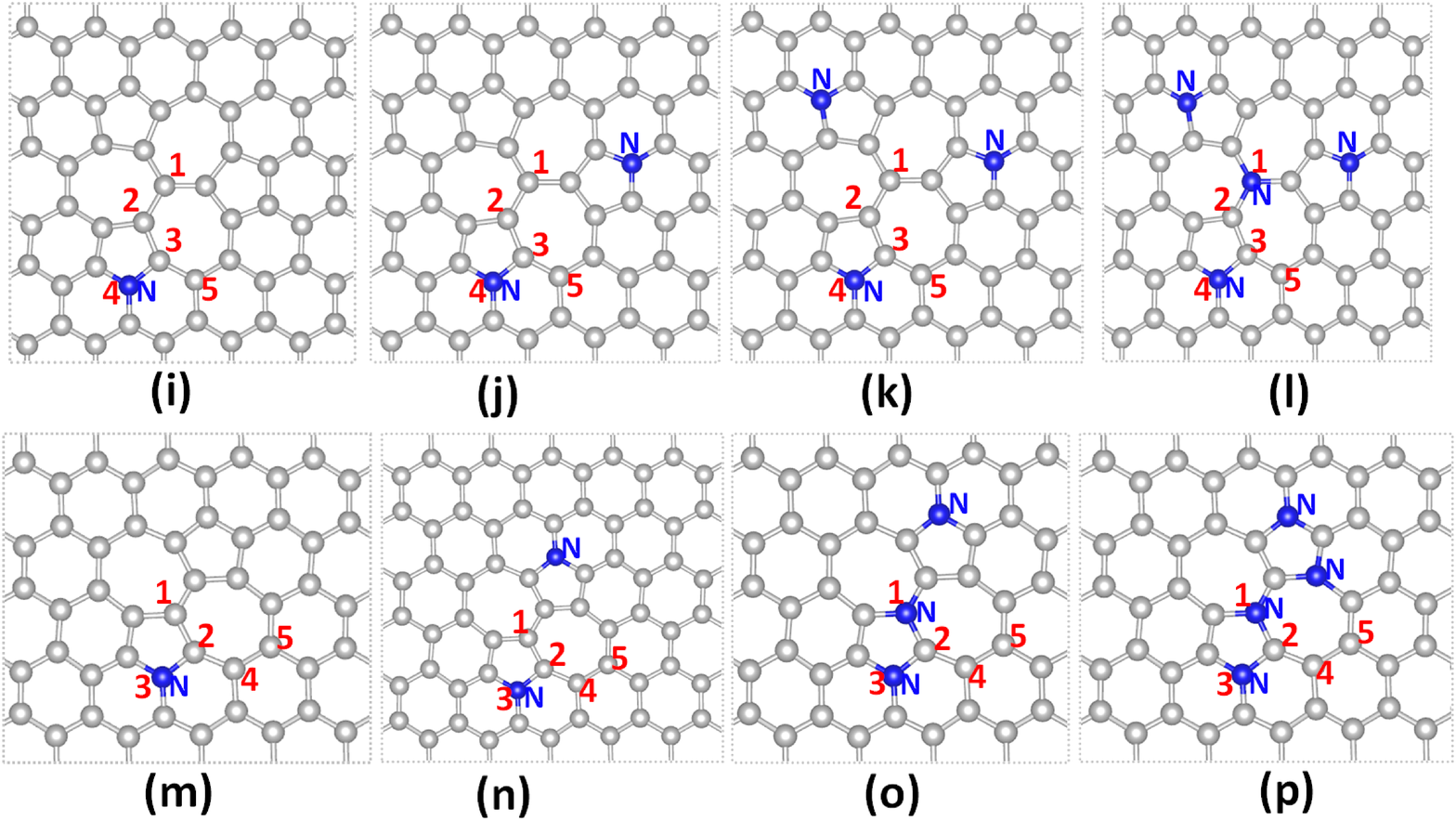}
\end{minipage}
\caption{\label{fig:4} (Color online) The optimized structures for the most stable configurations of \emph{m} N dopants ($m=1$, 2, 3, and 4) in defective graphene with MV [(a)-(d)], 5-8-5 DV [(e)-(h)], 555-777 DV [(i)-(l)], and SW defect [(m)-(p)]. The blue and light gray balls stand for the N and C atoms, respectively. }
\end{center}
\end{figure*}

For defect-free graphene, as the interaction between two N dopants is repulsive [Fig.~\ref{fig:2}(a)], aggregation of N dopants in a small region is unlikely.  However, we showed that a N dopant and a NPD attract each other quite strongly. Therefore, we expect that NPD may induce N dopant aggregation. Figure~\ref{fig:4} presents the most stable configurations of multiple N substitutions ($m=$ 2, 3, and 4) in defective graphene.  First we point out some characteristic features seen in the stable configurations. In the presence of MV, one of two N dopants occupies the site next to the vacancy to form a pyridinelike N and the other one substitutes for a C atom at a five-membered ring [see Fig.~\ref{fig:4}(b)]. This configuration is stable by 0.257 eV compared to that of two pyridinelike N atoms at a MV. For three N dopants at MV, all of them form the pyridinelike configuration. Such pyridinelike N at MV in carbon nanotube (CNT) is proposed to be responsible for the introduction of a large electron donor state in N-doped CNT.~\cite{Czerw2001} The present study shows that the configuration of the three pyridinelike N atoms at MV has a very high stability against the others. For example, the total energy difference between the most stable configuration [see Fig.~\ref{fig:4}(c)] and the second stable one [N$_\mathrm{C1}+\mathrm{N_{C7}}+\mathrm{N_{C8}}$, where C8 is the second nearest neighbor of C1 as shown in Fig.\ref{fig:4}(c).] of three N substitutions at MV is about 1.763 eV. For two N dopants at the 5-8-5 DV the most stable configuration corresponds to two pyridinelike N atoms [see Fig.~\ref{fig:4}(f)]. This configuration is more stable than the one with two N dopants at C4 sites by 0.223 eV. These two pyridinelike nitrogens tend to be closer together, contrary to the corresponding nitrogens at the armchair edge of GNR.~\cite{Jain2006}  Up to three N dopants at the 555-777 DV or up to two N dopants at the SW defect, the most stable configuration is the combined single N dopants, where substitutional N atoms occupy the vertex sites in different five-membered rings.

\begin{figure*}[htbp!]
\begin{center}
\begin{minipage}[t]{8.0cm}
\includegraphics*[width=8.0cm]{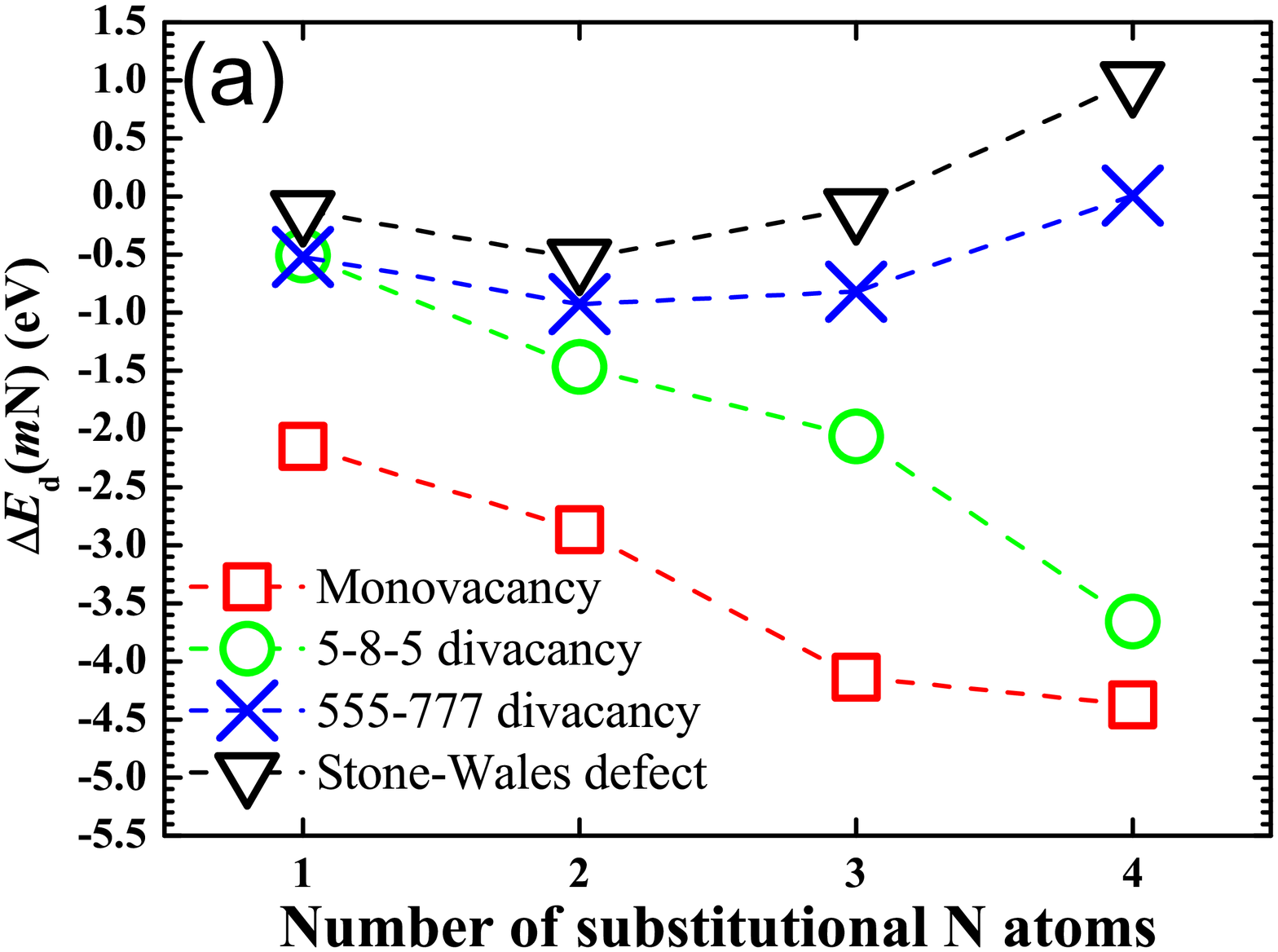}
\end{minipage}
\begin{minipage}[t]{8.0cm}
\includegraphics*[width=8.0cm]{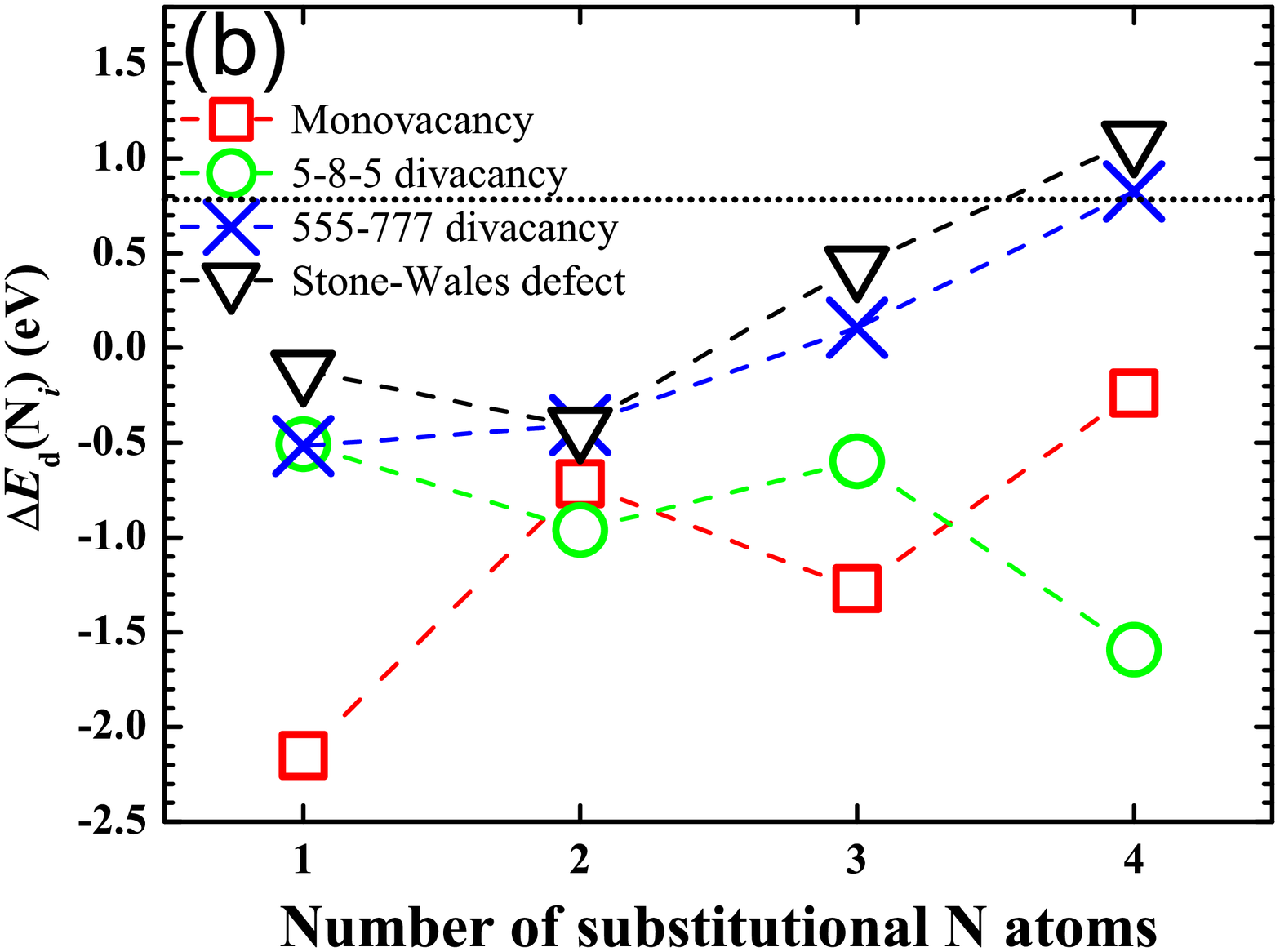}
\end{minipage}
\caption{\label{fig:5} (Color online) (a) The formation energy of N dopants [$\Delta E_\mathrm{d}(m\mathrm{N})$] in graphene with a NPD of some different kinds as a function of the number of substitutional N atoms. (b) The formation energy of \emph{i}th N dopant in a system
containing already ($i-1$) N dopants near a NPD [$\Delta E_\mathrm{d}(\mathrm{N}_i)$] for $i$ in the range from 1 to 4. The formation energy of single N dopant in perfect graphene is indicated with the dotted line in (b).}
\end{center}
\end{figure*}

Now we discuss the possibility of N dopant aggregation near a NPD in a quantitative way. We generalize Eq.~(\ref{eq:4-5}) to the case of multiple N dopants where $m$ N atoms substitute for $m$ C atoms near a NPD to obtain
\begin{equation}\label{eq:9}
 \Delta E_\mathrm{d}(m\mathrm{N})=E_{m\mathrm{N+d}}-E_\mathrm{d}+m\mu_\mathrm{C}-m\mu_\mathrm{N}.
\end{equation}
Figure~\ref{fig:5}(a) presents $\Delta E_\mathrm{d}(m\mathrm{N})$ for the configurations in Fig.~\ref{fig:4}. It is clear that N dopants are particularly stable at MV and that four N dopants at 5-8-5 DV shows also strong stability.  The formation energy of a complex of $m$N dopants plus a NPD is given by the sum of the NPD formation energy $\Delta E_\mathrm{0}(\mathrm{d})$ of Table~\ref{tab:1} and $\Delta E_\mathrm{d}(m\mathrm{N})$.  Although the formation energy of a NPD without N doping is smallest (largest) for SW (MV), 3N doping at the NPD makes MV have the smallest formation energy.  Furthermore, the formation energy of 5-8-5 DV, which is the second largest in undoped graphene, sharply decreases with N doping and becomes the second smallest with 4N doping.

In addition to these quantities, we define the formation energy of $i$th N dopant in a system containing already $(i-1)$ N dopants near a NPD by the following equation:
\begin{equation}\label{eq:10}
 \Delta E_\mathrm{d}(\mathrm{N}_i)=E_{i\mathrm{N+d}}-E_{(i-1)\mathrm{N+d}}+\mu_\mathrm{C}-\mu_\mathrm{N}.
\end{equation}
Clearly, $\Delta E_\mathrm{d}(\mathrm{N}_i)$ is the finite difference of $\Delta E_\mathrm{d}(i\mathrm{N})$ with respect to $i$ and the following relation holds.
\begin{equation}\label{eq:11}
\Delta E_\mathrm{d}(m\mathrm{N})=\sum\limits_{i=1}^m \Delta E_\mathrm{d}(\mathrm{N}_i).
\end{equation}
Figure~\ref{fig:5}(b) shows $\Delta E_\mathrm{d}(\mathrm{N}_i)$ for $i$ ranging from one to four for each of the different NPDs.  So long as $\Delta E_\mathrm{d}(\mathrm{N}_i)$ is smaller than 0.785 eV, which is the formation energy of N dopant in perfect graphene indicated with the dotted line in Fig.~\ref{fig:5}(b), an aggregate of $i$ N dopants at the NPD is energetically stable if other NPDs are not present.

However, the situation may be different in the presence of other NPDs.  We consider the case where two NPDs, designated as d$_1$ and d$_2$, exist before N doping and calculate the following energy
\begin{equation}\label{eq:12}
E(k_1\mathrm{N+d}_1, k_2\mathrm{N+d}_2)=\Delta E_{\mathrm{d}_1} (k_1\mathrm{N})+\Delta E_{\mathrm{d}_2}(k_2\mathrm{N}),
\end{equation}
which is the sum of formation energies of $k_1$N dopants at a d$_1$-NPD and $k_2$N dopants at a d$_2$-NPD.  We then find $k_1$ $(k_2)$ which minimizes the above energy for a given set of d$_1$, d$_2$ and $m=k_1+k_2$.  We only present some examples of this analysis in the following.

First, we consider the following cases of equivalent two NPDs and present the calculated values of $E(k_1\mathrm{N+d}_1, k_2\mathrm{N+d}_2)$ in Table~\ref{tab:2}.

\begin{table*}[htbp]
  \caption{The sum of formation energies [$E(k_1\mathrm{N+d}_1, k_2\mathrm{N+d}_2)$, in eV] of $k_1$N dopants at a $\mathrm{d}_1$-NPD and $k_2$N dopants at a
$\mathrm{d}_2$-NPD for the case of $\mathrm{d}_1$ = $\mathrm{d}_2$. }
  \label{tab:2}
  \begin{tabular}{c cc ccc }
    \hline
    $\mathrm{d}_1$ = $\mathrm{d}_2$  & $m= k_1+k_2$  &  \multicolumn{4}{c}{ $E(k_1\mathrm{N+d}_1, k_2\mathrm{N+d}_2)$}\\
      \cline{3-6}
    &               &  $k_1=1$  & $k_1=2$  &$k_1=3$  &  $k_1=4$ \\
    \hline
     MV &  2 & -4.298 & -2.865 & $-$ &  $-$\\
     MV &  3 & -5.015 & -5.015 &-4.136  &  $-$    \\
     MV &  4 & -6.285 & -5.731 & -6.285 & -4.373      \\
     5-8-5 DV & 2 & -1.012 & -1.467 & $-$ &  $-$    \\
     5-8-5 DV & 3 & -1.973& -1.973 &-2.065 &  $-$    \\
     5-8-5 DV  & 4 & -2.572& -2.934& -2.572 &-3.657     \\
     555-777 DV & 2 & -1.035& -0.924 & $-$ &  $-$   \\
     555-777 DV & 3 & -1.442& -1.442 &-0.818   &  $-$   \\
     555-777 DV  & 4 & -1.336  &  -1.849 & -1.336 & 0.008     \\
     SW &  2 & -0.237  & -0.535 &   $-$ &  $-$   \\
     SW & 3 & -0.654  & -0.654 &-0.101 &   $-$   \\
     SW & 4 & -0.220  & -1.071 & -0.220 &0.993   \\
    \hline
  \end{tabular}
\end{table*}

(1) d$_1$ = d$_2$ = SW: For $m$ = 2, ($k_1=2$, $k_2=0$) or equivalently ($k_1=0$, $k_2=2$) is more stable than  ($k_1=1$, $k_2=1$).  For $m$=3, ($k_1=2$, $k_2=1$) or equivalently ($k_1=1$, $k_2=2$) is more stable than  ($k_1=3$, $k_2=0$) and ($k_1=0$, $k_2=3$).   Therefore aggregation of 2 N dopants is possible.

(2) d$_1$ = d$_2$ = 5-8-5 DV: For $m$ = 2, the situation is the same as in SW. For $m = 3$, the situation is opposite to the one in SW. For $m=4$, ($k_1=4$, $k_2=0$) or equivalently ($k_1=0$, $k_2=4$) is more stable than other partitions of 4 N dopants. From this analysis, we conclude that aggregation of 4 N dopants at one 5-8-5 DV is possible even if more 5-8-5 DVs may exist before N doping.

(3) d$_1$ = d$_2$ = MV: As the formation energy of the first N doping is strongly negative, for any $m>1$, ($k_1=1$, $k_2=m-1$) is not stable among other partitions.  Therefore, if many MV may exist, the configuration with the maximum number of MVs with single N dopant will be most stable.

For inequivalent two NPDs (see Table S1 in the Supplemental Material at the link in \onlinecite{suppm}), if one of the two NPDs (d$_1$, for example) is MV, selective N doping will occur at MV due to the large negative formation energy of N dopants at MV.  However, in our analysis, if the other NPD is 5-8-5 DV or 555-777 DV and $m=4$, ($k_1=3$, $k_2=1$) is slightly more stable than ($k_1=4$, $k_2=0$).

\begin{table*}[htbp]
  \caption{FeN$_x$ centers are formed by the Fe adsorbate and the pyridinelike N at MV and 5-8-5 DV in graphene: adsorption energy [$E_\mathrm{ad}$(Fe), in eV/atom] of Fe atom, total magnetic moment ($M_\mathrm{tot}$, in $\mu$B), local magnetic moment of Fe atom ($m_\mathrm{Fe}$, in $\mu$B/atom), bond length between Fe and its nearest neighbors [$d_\mathrm{Fe-N}$ or $d_\mathrm{Fe-C}$, in \AA], and height of Fe atom ($z_\mathrm{Fe}$, in \AA) with respect to the atomic plane of graphene sheet. The adsorption energy of Fe atom is calculated as follows: $E_\mathrm{ad}$(Fe) = $E_{\mathrm{Fe}+m\mathrm{N+d}} -E_\mathrm{d}-\mu_\mathrm{Fe}$, where $E_{\mathrm{Fe}+m\mathrm{N+d}}$ is the total energy of  FeN$_x$ center embedded in graphene. $\mu_\mathrm{Fe}$ is the chemical potential potential of iron and  here it is taken as the total energy of a Fe atom. }
  \label{tab:3}
  \begin{tabular}{cc cc ccc}
    \hline
  Configuration & $E_\mathrm{ad}$(Fe)  & $M_\mathrm{tot}$ & $m_\mathrm{Fe}$ & $d_\mathrm{Fe-N}$& $d_\mathrm{Fe-C}$ & $z_\mathrm{Fe}$\\
    \hline
   1N at MV   & -5.425 & 0.96    & 0.83   & 1.744 & 1.763   & 1.574 \\
   3N at MV   & -4.837 & 3.40   &  3.11   & 1.880  & -          & 1.789 \\
  2N at 5-8-5 DV & -7.230  & 2.30  &  2.70  & 1.959  &  1.879  &  0.0  \\
  4N at 5-8-5 DV & -7.579  & 1.98  &  2.06   &  1.895   & -    & 0.0\\
    \hline
  \end{tabular}
\end{table*}

Finally we discuss the effect of the adsorbed Fe atom on the stabilities of the \emph{m} pyridinelike N dopants at the MV [$m=1$ and 3, as shown in Fig.~\ref{fig:4}(a) and (c)] and 5-8-5 DV [$m=2$ and 4, as shown in Fig.~\ref{fig:4}(f) and (h)].
Table ~\ref{tab:3} is a summary of calculated results.  Though detailed discussion on the physical and chemical properties of Fe adsorption to N-doped graphene will be given separately, we here point out an important consequence of Fe adsorption on the stability of N dopant configuration. Combining $\Delta E_0(\mathrm{d})$ of Table~\ref{tab:1}, $\Delta E_{\mathrm{d}}(m\mathrm{N})$ in Fig.~\ref{fig:5}(a) and $E_{\mathrm{ad}}$(Fe) in Table ~\ref{tab:3}, we find that
the formation energy~\cite{formfe} of four N dopants at 5-8-5 DV becomes lower than that of three N dopants at the MV by 2.357 eV in the presence of adsorbed Fe, while the former is higher than the latter by 0.385 eV in the absence of adsorbed Fe. This suggests that four pyridinelike N dopants at the 5-8-5 DV may be quite stable in the presence of iron.  The strong stability of the FeN$_4$ complex was pointed out also in the carbon nanotube.~\cite{DHLee2011}

\section{Conclusion}
\label{sec:4}
We have studied the effect of native point defects on the N doping of graphene using DFT electronic structure calculations. Our calculations show that substitutional N doping tends to occur at the carbon sites with larger shrinkage of the defect-induced bond and also at the vertex site of a pentagonal ring if it exists. The presence of native point defects can lower the formation energy of N dopant from being endothermic in defect-free graphene to being exothermic in defective graphene. This suggests that the intentional creation of defect before introducing N dopant will enhance the N doping of graphene. On the other hand, the formation energy of defects is reduced after N doping, indicating that N doping would increase the probability of point defect generation in graphene.  We also analyzed possibilities of multiple N doping at a NPD due to the strong attractive interaction between a native point defect and a N dopant. In the actual N doping process of graphene, the partial pressure of N$_2$ or NH$_3$ gas, the high-temperature heat treatment, and the kinetic factors may also affect the distribution of doped N.  Nevertheless, our analysis may give some insight into the stability of N dopant configuration in the defective graphene.  In the present paper, we focused on structural and energetics aspects of the problem.  The detailed discussion on the underlying electronic structures will be given in a future publication.

\section*{Acknowledgement}
This work was performed under Project 08003441-0 at the New Energy and Industrial Technology Development Organization (NEDO).
The computation was performed using the supercomputing facilities in the Center for Information Science in JAIST. Parts of the computations were done on TSUBAME Grid Cluster at the Global Scientific Information and Computing Center of the Tokyo Institute of Technology.
\bibliography{reference}

\begin{thebibliography}{57}
\bibitem[{for()}]{formfe}
\bibinfo{note}{The formation energies of the Fe adsorbed on the $m$
  pyridine-like N dopants at the MV and 5-8-5 DV are calculated as follows:
  $\Delta E_{\mathrm{Fe}+m\mathrm{N+d}} = E_{\mathrm{Fe}+m\mathrm{N+d}}- E_0
  +(m+n_\mathrm{C}) \mu_\mathrm{C}-m \mu_\mathrm{N}-\mu_\mathrm{Fe}$.}
\end{thebibliography}

\end{document}